\documentclass[acmsmall]{acmart}
\usepackage{graphicx} 
\usepackage{subfig}
\usepackage[utf8]{inputenc}
\usepackage{xcolor}
\usepackage[inline]{enumitem}
\usepackage{mdframed}
\usepackage{comment}
\usepackage{booktabs}
\usepackage{longtable}
\usepackage{flushend}
\usepackage{multirow}
\usepackage{comment}
\newmdenv[skipabove=6mm]{kotak}   
\usepackage{float}
\usepackage{multirow}
\setlist[enumerate,1]{label=\textit{\alph*)}}
\usepackage{amsmath,amsfonts}
\usepackage{textcomp}
\usepackage{hyperref}
\usepackage{afterpage}
\usepackage{rotating}
\usepackage{arydshln}
\usepackage{listings}
\usepackage{xcolor}

\usepackage{todonotes}
\usepackage[inline]{enumitem}
\colorlet{punct}{red!60!black}
\definecolor{background}{HTML}{EEEEEE}
\definecolor{delim}{RGB}{20,105,176}
\colorlet{numb}{magenta!60!black}
\lstdefinelanguage{json}{
    basicstyle=\normalfont\ttfamily,
    numbers=left,
    numberstyle=\scriptsize,
    stepnumber=1,
    numbersep=8pt,
    showstringspaces=false,
    breaklines=true,
    frame=lines,
    backgroundcolor=\color{background},
    literate=
     *{0}{{{\color{numb}0}}}{1}
      {1}{{{\color{numb}1}}}{1}
      {2}{{{\color{numb}2}}}{1}
      {3}{{{\color{numb}3}}}{1}
      {4}{{{\color{numb}4}}}{1}
      {5}{{{\color{numb}5}}}{1}
      {6}{{{\color{numb}6}}}{1}
      {7}{{{\color{numb}7}}}{1}
      {8}{{{\color{numb}8}}}{1}
      {9}{{{\color{numb}9}}}{1}
      {:}{{{\color{punct}{:}}}}{1}
      {,}{{{\color{punct}{,}}}}{1}
      {\{}{{{\color{delim}{\{}}}}{1}
      {\}}{{{\color{delim}{\}}}}}{1}
      {[}{{{\color{delim}{[}}}}{1}
      {]}{{{\color{delim}{]}}}}{1},
}

\def\BibTeX{{\rm B\kern-.05em{\sc i\kern-.025em b}\kern-.08em
    T\kern-.1667em\lower.7ex\hbox{E}\kern-.125emX}}

\newcommand{\Rbb}{\mathbb{R}}

\newcommand{\A}{\mathbf{A}}
\newcommand{\B}{\mathbf{B}}
\newcommand{\Cbf}{\mathbf{C}}

\newcommand{\F}{\mathbf{F}}
\newcommand{\Ft}{\tilde{\F}}

\newcommand{\Y}{\mathbf{Y}}
\newcommand{\Z}{\mathbf{Z}}

\AtBeginDocument{%
  \providecommand\BibTeX{{%
    Bib\TeX}}}

\begin{document}

\title{Identifying and Explaining Safety-critical Scenarios for Autonomous Vehicles via Key Features}


\author{Neelofar Neelofar}
\affiliation{%
  \institution{Monash University}
  \city{Melbourne}
  \country{Australia}}
\email{neelofar.neelofar@monash.edu}

\author{Aldeida Aleti}
\affiliation{%
 \institution{Monash University}
 \city{Melbourne}
  \country{Australia}}
  \email{aldeida.aleti@monash.edu}


\begin{abstract}\label{abstract}
Ensuring the safety of autonomous vehicles (AVs) is of utmost importance, and testing them in simulated environments is a safer option than conducting in-field operational tests. However, generating an exhaustive test suite to identify critical test scenarios is computationally expensive as the representation of each test is complex and contains various dynamic and static features, such as the AV under test, road participants (vehicles, pedestrians, and static obstacles), environmental factors (weather and light), and the road's structural features (lanes, turns, road speed, etc.). In this paper, we present a systematic technique that uses \textit{Instance Space Analysis (ISA)} to identify the significant features of test scenarios that affect their ability to reveal the unsafe behaviour of AVs. ISA identifies the features that best differentiate safety-critical scenarios from normal driving and visualises the impact of these features on test scenario outcomes (safe/unsafe) in $2D$. This visualisation helps to identify untested regions of the instance space and provides an indicator of the quality of the test suite in terms of the percentage of feature space covered by testing. To test the predictive ability of the identified features, we train five Machine Learning classifiers to classify test scenarios as safe or unsafe. The high precision, recall, and F1 scores indicate that our proposed approach is effective in predicting the outcome of a test scenario without executing it and can be used for test generation, selection, and prioritisation.
\end{abstract}

\maketitle

\section{Introduction}\label{sec:introduction}
Autonomous vehicles (AVs) are expected to operate in an ever-changing environment with a large and complex space of scenarios. This requires extensive testing of these vehicles to ensure their safety and reliability before they are deployed in the real world. Assessing the reliability of AVs requires hundreds of millions of miles of test-driving~\cite{kalra2016many}. This level of testing is not possible using on-road driving alone. Furthermore, on-road testing cannot be performed for safety-critical driving scenarios. Therefore, the use of simulated environments for the testing of AVs is a norm~\cite{kim2016testing,zofka2016testing,math2013opends}. 
Just like on-road testing, testing in a simulated environment requires designing a huge number of driving scenarios, thus attracting developers and testers to devise automated testing techniques for AV-testing~\cite{rajabli2020software, hauer2019fitness}. To identify the safety violations of AVs, evolutionary algorithms are applied to generate safety-critical test scenarios~\cite{abdessalem2018testing, ben2016testing, kluck2019genetic, li2020av, lu2021search, onieva2015multi, tian2022mosat}. However, it is a challenging task, as the operating environment is dynamic, with an exceedingly large number of factors that impact driving. These factors include, but are not limited to: the kinematics of AV and other participants sharing the road (vehicles, humans and static obstacles), structural features of roads (lanes, speed limit, curve, pedestrian crossing) and environmental factors like weather (rain, fog) and light conditions. To test the system for reliability and safety, it is crucial to consider all of these factors for test generation. However, the number of features representing these factors is huge. For instance, the representation of an NPC in a simulated scenario requires defining its features like starting position, destination, speed, type of vehicle, action, the time of action etc., and a single scenario can have many of such vehicles. Similarly, pedestrians, weather and other factors have their own features. Defining the search space based on all these features would fuel the complexity of the search space, making it infeasible to explore completely. One way of curbing the complexity of the search space is by gaining deeper insights into the impact of the features on the effectiveness of the test scenarios (i.e., their ability to detect incorrect behaviour of the AV), and searching the space for these impactful features only. 

Reducing the cost of AV testing is an active research area in the software testing community, and researchers have proposed frameworks and techniques to minimise the number of test scenarios using test selection and prioritisation strategies~\cite{birchler2021automated, birchler2022single, lu2021search}. However, the main goal of these studies is the selection of \textit{a subset of test-cases} from a test suite. Unlike these studies, we propose to curb the complexity of the search space of testing by selecting a \textit{subset of features} to define a simulated driving scenario, therefore, reducing the curse of dimensionality. The selected set of features can effectively be used for test generation, selection and prioritisation. In short, the test selection and prioritisation studies mentioned above decrease the size of a test suite in terms of its length (number of test cases), while we propose to shrink this size in width (number of features to define a test case). Furthermore, we propose to project the features and the test outcomes in $2D$ space in such a way that the impact of each feature, along with its value range, can easily be mapped to the test outcome, making the feature selection more explainable and human-interpretable. This will facilitate the developers and testers to understand the structural and behavioural factors that impact the safety of the AV under test.

To gain the necessary insights into the impact of features on AV testing, we investigate the following research questions:

\textbf{RQ1}: How can we identify key features of effective test scenarios? Generating safety-critical test scenarios using search-based techniques is an active research area in the field of automated testing of AVs. In these techniques, the search is guided by a fitness function that evaluates the testing scenarios based on metrics such as time-to-collision~\cite{minderhoud2001extended}, Out of Bound Episodes (OBEs)~\cite{birchler2021automated}, collision-probability~\cite{lu2022learning} etc. We call these metrics \textit{test output}, and test scenarios need to be executed in the simulator for their computation. However, it is still rarely investigated what features should be considered to design a test scenario. These features are known as test parameters or input features. In this research question, we find the key input features that impact the effectiveness of test scenarios. Effectiveness is defined as the ability of the scenarios to detect faults in the system. For this, we employ \textit{Instance Space Analysis} (ISA)~\cite{kang2017visualising, munoz2017performance, smith2014towards}, a framework that constructs a two-dimensional instance space based on the input features of test scenarios and their outcome. The generated instance space provides visual insights into the impact of various features (test input) on the effectiveness of testing scenarios (test output). 

\textbf{RQ2}: Can we predict test scenario outcomes using the key features? The simulators provide a safer substitute for in-field operational testing of AVs~\cite{afzal2021simulation, timperley2018crashing, gambi2019generating, lou2022testing}. However, the testing space is huge, and the tests are expensive to run even on a simulator. Using the key features investigated in RQ1, we train machine learning classifiers to predict the effectiveness of a test scenario without simulating or executing it. Obtaining such information in advance would make the testing process more effective and faster by prioritising the test scenarios for simulation which are more likely to lead to a collision.

\textbf{RQ3}: How can we measure the extent to which the scenario space has been explored? A crucial problem in testing is deciding when the system has been tested sufficiently. For the testing of traditional systems, code coverage is widely used as a measure of testing adequacy~\cite{harman2007theoretical, ghani2009comparing, scalabrino2016search, panichella2017lips, panichella2018large, campos2017empirical, wu2018empirical}. However, this measure cannot be used as a quality metric for the simulation-based testing of AVs, as the system logic is not completely encoded in the source code but is largely shaped by the training data. Therefore, a testing criterion specific to this testing problem is needed. One way of making sure that testing has exercised diverse behaviours of the system-under-test is to explore the scenario space at large. As driving scenarios for AV-testing are defined in terms of input features, it is safe to use the terms \textit{feature space} and \textit{scenario space} interchangeably. Under this research question, we seek a metric to assess the effectiveness of AV testing by measuring how extensively the scenario space has been explored. 

To answer these questions, we extract a set of features from three AV testing suites for which test outcomes are known. These features include the structural properties of the road, the properties of the AV, pedestrians, other vehicles on the road, traffic signals, and weather and light conditions. The features are then used to generate the instance space that reveals the impact of these features on scenario outcome. The scenario outcome is labelled as unsafe or safe depending on its execution result. The scenario is deemed \textit{unsafe} if it drifts away from the centre of the lane, leads to a collision or in which AV comes very close to the other road participants and has a chance to collide with obstacles, \textit{safe} otherwise.

We extract a subset of features that have maximum impact on the testing scenarios to generate the instance space that provides visual insights into how the features impact scenario outcome. Based on the minimum and maximum values of the features, we define an empirical boundary around the generated instance space and identify the regions where more test scenarios should be added. Finally, to evaluate the effectiveness of the features selected through our proposed technique, we train four machine learning classifiers using these features to predict test outcomes. The performance of the trained classifiers, measured in terms of precision, recall, and F1 score, is significantly higher than random feature selection.

The approach introduced in this paper can be used by the testers to generate effective test scenarios using a smaller number of features. This can help to reduce the testing effort in terms of time and computational resources without compromising the performance of the test suite. The identification of the percentage of missing scenarios provides an objective measure to assess the adequacy and sufficiency of testing, whereas, the prediction of the outcome of a test scenario without executing it on a simulator can be a valuable technique for test-case selection and prioritisation.

\section{Instance Space Analysis} \label{sec:ISA}

ISA is a methodology that generates a $2D$ visualisation of the problem instances and shows the impact of the features of a problem on the performance. In the context of testing, we use this methodology to identify the impact of input features of test scenarios on their outcome (safe vs. unsafe), and project these scenarios in the $2D$ space, called instance space, in such a way that this impact can be clearly visualised. The generated instance space facilitates insights into the
distribution of existing instances, allowing the identification of sparse, unoccupied regions where new test instances should be generated for more comprehensive testing. \color{black} The ISA methodology was initially proposed by Smith-Miles et al.~\cite{smith2014towards} for combinatorial optimisation problems but has since been applied to a multitude of other areas, including automated software testing~\cite{oliveira2018mapping, smith2022instance} and automated program repair~\cite{aleti2020apr}. In general, ISA involves the following steps. 
\begin{itemize}
    \item Collecting the metadata for test instances: features of test instances and their performance based on a performance metric.
    \item Selecting a subset of features that best define the performance and capture differences and similarities in the problem instances.
    \item Projecting test instances from n-dimensional feature space to a 2-dimensional instance space. 
    \item Quantitatively measuring the regions of good performance in the instance space (called footprint) and describing them in terms of instance features.
    \item Defining an empirical boundary around the instance space that encloses all the test scenarios that are empirically possible to generate, however, may be missing from the current analysis.  
    \item Constructing an automated recommender that predicts scenario outcome without executing a test scenario in a simulator. 
\end{itemize}  



\begin{figure}[!ht]
    \centering
    \includegraphics[width=0.8\linewidth]{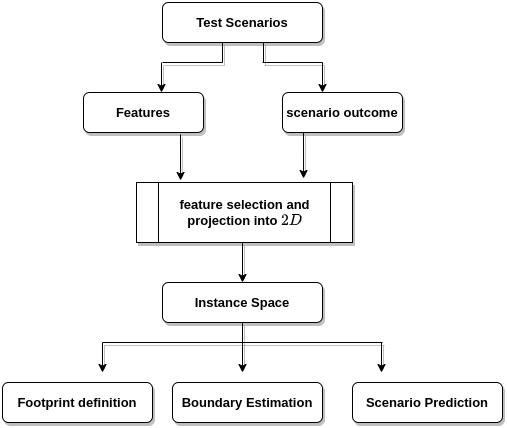}
    \caption{Instance Space Analysis Framework}
    \label{fig:ISA}
\end{figure}

\subsection{ISA for testing AVs}
In the context of creating the instance space for testing AVs, ISA begins with the compilation of a wide variety of test scenarios. These scenarios must be diverse both in terms of feature values and their execution results (fail vs. pass). The test execution result is computed based on a performance metric. In the context of AV testing, some examples of performance metrics are: collision probability~\cite{lu2022learning}, collision event (collision or no collision)~\cite{lu2021search}, Out of Bound Episodes (OOBs)~\cite{birchler2021automated}, and Time to Collision (TTC)~\cite{kluck2019genetic}. 

In the next step, meaningful features are extracted from the test scenarios such that they: \begin{enumerate*}
    \item can describe the similarities and differences between instances; \item are useful in explaining the effectiveness of the test scenarios; and \item can be computed within a reasonable amount of time.
\end{enumerate*}
Features are domain-specific, and feature extraction requires significant domain knowledge~\cite{munoz2018instance,mario2020regression}. We list the features of the test scenarios we use in our study and their extraction process in Section~\ref{sec:feat_ext}.

Ultimately, we construct an instance space, representing test scenarios in a $2D$ format, focusing on features with the greatest impact on scenario outcomes. The test scenarios are projected onto this $2D$ space in a manner that reflects a gradual progression of feature distributions from one end of the instance space to the other. This projection allows for a clear mapping of scenario outcomes, which are also projected in a similar manner i.e., the safe scenarios occupy one end of the space, while unsafe scenarios reside at the opposite end.

Once the instance space is established, it becomes a valuable tool for assessing the quality of the test suite used to generate it. Additionally, the features identified by the instance space are utilised to train machine learning algorithms, which are subsequently employed to predict the outcomes of untested scenarios without the need for their execution in simulators.

\subsection{Features of Driving Scenarios}\label{sec:feat_ext}

To generate the instance space of AV testing, we utilise two different test suites having test scenarios designed for testing AVs from different perspectives. The first test suite is designed for the system-level testing of AVs and the test scenarios are defined by the combination of 
\begin{enumerate*}
    \item static features to represent road and weather conditions;
    \item dynamic features representing the state and behaviour of AV and other road users and defining their interaction.
\end{enumerate*}

Static road features determine the background of a scenario. They include features to characterise the road, e.g. the state of the traffic light, the presence of sidewalks, the number of AVs, pedestrians and other obstacles on the road etc. Additionally, within the static features' category, we account for weather and lighting conditions as these factors can significantly influence sensor data quality. For instance, adverse weather conditions like heavy rain can lead to blurred sensor images, potentially resulting in challenges for AV's performance. 

The dynamic features capture the spatio-temporal data of AV and other road participants. Along with storing the absolute values of position, speed, velocity, acceleration, volume and operation of the road users, we include the relative values of these features. For instance, $vol\_maxSpeedObs$ represents the volume of the fastest obstacle (vehicle or pedestrian) on the road and $speed\_minDistObs$ stores the speed of the obstacle that is in the closest vicinity of the AV. 

We extract a diverse set of 61 static and dynamic features from this test suite, listed in Table~\ref{tab:feat_ds1}. The feature set not only incorporates the features utilised in the prior studies but also encompasses a considerably larger number of additional features~\cite{lu2021search, lu2022learning, hauer2020clustering, abdessalem2018testing, deng2022scenario}. A sample representation of the driving scenario from this test suite is provided in Appendix~\ref{sec:appendixa}. 

The second test suite we use in our study is specifically designed to evaluate the lane-keeping capabilities of autonomous vehicles (AVs). Within this suite, the test scenarios incorporate various structural features, which serve as the basis for creating virtual roads within a simulation environment. The primary objective of this test suite is to create complex road conditions that challenge AVs, potentially causing them to deviate from the centre of their designated lanes, leading to unsafe situations. From these test scenarios, we derive 15 distinct features that encompass information such as the frequency of left and right turns, the curvature of the turns, the angularity of the turns, the number of straight road segments, the overall road length, and summarised statistics of these features, including minimum, maximum, standard deviation, mean, and median values. These features have already been used in previous studies related to the testing of AVs~\cite{birchler2022cost, birchler2022single, birchler2021automated}. They are listed in Table~\ref{tab:feat_dS2} and a sample test scenario from this test suite is provided as Appendix~\ref{sec:appendixb}. 

\afterpage{
\begin{longtable}{p{0.05\linewidth}|p{0.27\linewidth} p{0.57\linewidth}} 
     \caption{Features of Test Scenarios: Dataset1.\label{tab:feat_ds1}}\\
    \toprule
       
    \multirow{6}{*}{\begin{turn}{-90} AV\end{turn}} & F1: AV\_acceleration & initial acceleration of AV defined on x, y and z axis\\
    & F2: AV\_throttle & initial throttle of AV\\
    & F3: AV\_brake & initial braking force applied by AV\\
    & F4: AV\_steeringRate & initially defined steering rate of AV\\
    & F5: AV\_speed & initial speed of AV\\
    & F6: AV\_velocity & initial velocity of AV defined on x and y axis\\
    & F7: AV\_position & initial position of AV defined on x, y and z axis\\\\
    \cdashline{1-3}
    \\
    
    \multirow{7}{*}{\begin{turn}{-90} road and traffic\end{turn}} & F8: num\_peds & number of pedestrians in a scenario\\
    & F9: num\_NPCs & number of non-player characters (other vehicles) participating in a scenario\\
    & F10: num\_statObs & number of static obstacles included in a scenario\\
    & F11: tot\_obs &   total number of obstacles in a scenario\\
    & F12: is\_ped & boolean value showing the presence/absence of pedestrians in a multi-obstacle scenario\\
    & F13: traffic\_light & traffic signal (red, orange, green or no signal)\\
    & F14: Sidewalk & boolean value showing presence/absence of sidewalks in a scenario\\\\
        \cdashline{1-3}
    \\
  
    \multirow{4}{*}{\begin{turn}{-90} weather and light\end{turn}} & F15: rain & intensity of rain\\
    & f16: fog & intensity of fog\\
    & f17: wetness & intensity of wetness\\
    & f18: time of day & morning, noon or night time as a representative of light conditions\\\\
        \cdashline{1-3}
    \\
  
    \multirow{7}{*}{\begin{turn}{-90} dynamic participants\end{turn}} & F19: avg\_obsDist		&	average distance between AV and other obstacles	\\
    & F20: max\_obsDist		&	distance between AV and the obstacle at the maximum distance\\
    & F21: min\_obsDist		&	distance between AV and the obstacle at the minimum distance\\
    & F22: type\_maxDistObs		&	type of the obstacle at the maximum distance\\
    & F23: type\_minDistObs		&	type of the obstacle at the minimum distance 		\\
    & F24: op\_maxDistObs		&	operation performed by the obstacle at a maximum distance \\
    & F25: op\_minDistObs		&	operation performed by the obstacle at a minimum distance		\\
    & F26: vol\_minDistObs & volume of the obstacle at the minimum distance\\
    & F27: speed\_minDistObs & speed of the obstacle at the minimum distance\\
    & F28: dist\_minDistObs & actual distance of the closest obstacle to AV\\
    & F29: avg\_obsSpeed		&		average speed of the obstacles on the road	\\
    & F30: max\_obsSpeed		&		maximum speed of the obstacles on the road	\\
    & F31: min\_obsSpeed		&		minimum speed of the obstacles on the road	\\
    & F32: type\_maxSpeedObs		&	type of the obstacle having maximum speed		\\
    & F33: type\_minSpeedObs		&	type of the obstacle having minimum speed		\\
    & F34: op\_maxSpeedObs		&	operation performed by the maximum speed obstacle\\
    & F35: op\_minSpeedObs		&	operation performed by the minimum speed obstacle\\
    & F36: vol\_minSpeedObs & volume of the fastest obstacle\\
    & F37: speed\_minSpeedObs & speed of the fastest obstacle\\
    & F38: dist\_maxSpeedObs & distance of the fastest obstacle to AV\\
    & F39: avg\_obsVel		&	average velocity of the obstacles on the road		\\
    & F40: max\_obsVel		&	maximum velocity of the obstacles on the road		\\
    & F41: min\_obsVel		&	minimum velocity of the obstacles on the road		\\
    & F42: type\_maxVelObs		&	type of the obstacle having maximum velocity		\\
    & F43: type\_minVelObs		&	type of the obstacle having minimum velocity		\\	
    & F44: op\_maxVelObs		&	operation performed by the obstacle with maximum velocity		\\
    & F45: op\_minVelObs		&	operation performed by the obstacle with minimum velocity\\
    & F46: vol\_minVelObs		&	volume of the obstacle having maximum velocity\\
    & F47: avg\_obsAcc		&		average acceleration of the obstacles on the road	\\
    & F48: min\_obsAcc		&		maximum acceleration of the obstacles on the road	\\
    & F49: max\_obsAcc		&		minimum acceleration of the obstacles on the road	\\
    & F50: type\_maxAccObs		&	type of the obstacle having maximum acceleration		\\
    & F51: type\_minAccObs		&	type of the obstacle having minimum acceleration		\\
    & F52: op\_maxAccObs		&	operation performed by the obstacle with maximum acceleration		\\
    & F53: op\_minAccObs		&	operation performed by the obstacle with minimum acceleration		\\
    & F54: vol\_maxAccObs		&		volume of the obstacle having maximum acceleration	\\
    & F55: avg\_obsVol		&		average volume of the obstacles on the road	\\
    & F56: max\_obsVol		&		maximum volume of the obstacles on the road	\\
    & F57: min\_obsVol		&		minimum volume of the obstacles on the road	\\
    & F58: type\_maxVolObs		&	type of the obstacle having maximum volume		\\
    & F59: type\_minVolObs		&	type of the obstacle having minimum volume		\\
    & F60: op\_maxVolObs		&	operation performed by the obstacle having maximum volume		\\
    & F61: op\_minVolObs		&	operation performed by the obstacle having minimum volume	\\
    \bottomrule
\end{longtable}}

\section{Experimental Design}\label{sec:design}

This section provides insights into the dataset utilised within this study and elucidates the procedures and methods we employ to apply the ISA generic methodology, as outlined in Section~\ref{sec:ISA}, to construct the instance space for testing AVs.

\subsection{Datasets}\label{sec:dataset}

We use two case studies related to AV testing for our experiments. The first case study is the testing of Baidu Apollo Open Platform 5.0 as the ADS~\cite{apollo}, using LGSVL simulator~\cite{rong2020lgsvl} and is taken from the replication package provided with~\cite{lu2021search}. The test scenarios are generated using four different roads from San Francisco map. The test scenarios in this test suite are designed to mimic varied driving situations like taking turns, changing lanes, stopping on signals, approaching pedestrians/obstacles, overtaking, and parking etc. Figure~\ref{fig:sources} shows the distribution of these scenarios in the instance space according to each operational category. In total, 28,946 unique test scenarios are used in the current study, containing 12,930 unsafe and 16,016 safe scenarios. Failures in this dataset are defined as collisions or safety distance violations~\cite{shalev2017formal}.

The second case study is the testing of the BeamNG.AI lane-keeping system: the driving agent shipped with the BeamNG.tech simulator~\cite{BeamNG}. Unlike the first case study, the test scenarios in this dataset are the representation of challenging virtual roads that will force the AV to drive out of or very close to the road lane boundaries. For this case study, we generated test scenarios using two different testing strategies, i.e., AmbieGen~\cite{humeniuk2022ambiegen} and frenetic~\cite{frenetic}, presented in the SBST tool competition~\cite{panichella2021sbst, gambi2022sbst}. AmbieGen leverages NSGA-II as a multi-objective search algorithm to generate diverse virtual roads that will force the driving agent to go out of the road. The road is represented as a sequence of points, defining the
road spine. Frenetic is also a genetic approach, however, unlike AmbieGen, it uses curvatures associated with smooth planar curves
to represent roads. The test suite generated by Frantic contains 1294 scenarios, while the test suite generated by AmbieGen contains 514 test scenarios. 

\begin{table}[t!]
 \renewcommand{\arraystretch}{1.1}
    \centering
    \caption{Structural Features of Virtual Roads: Dataset2}
    \label{tab:feat_dS2}
      \resizebox{\columnwidth}{!}{
    \begin{tabular}{ p{0.05\linewidth}|p{0.27\linewidth} p{0.57\linewidth}}
    \toprule
    \multirow{15}{*}{\begin{turn}{-90} road features\end{turn}} & F1: min\_angle & The minimum angle turned in road segment\\
    & F2: max\_angle & The maximum angle turned in road segment\\
    & F3: mean\_angle & The average angle turned in road segment\\
    & F4: median\_angle & Median of angle turned in road segment\\
    & F5: std\_angle & Standard deviation of angle turned in road segment\\
    & F6: total\_angle & Total cumulated angle of all turns\\
    & F7: min\_pivot\_off	& The minimum radius of all turns of the road\\
    & F8: max\_pivot\_off & The maximum radius of all turns of the road\\
    & F9: mean\_pivot\_off & The average radius of all turns of the road\\
    & F10: median\_pivot\_off & The median of the radius of all turns of the road\\
    & F11: std\_pivot\_off & Standard deviation of the radius of all turns of the road\\
    & F12: num\_l\_turns & Number of left turns\\
    & F13: num\_r\_turns & Number of right turns\\
    & F14: num\_straights & Number of straight road segments\\
    & F15: road\_distance & Total length of the road\\    
    \bottomrule
\end{tabular}}
\end{table}

\subsection{Finding Effective Features of Test Scenarios (RQ1)}
\label{sec:finding_effective_features}
A critical step for the generation of an instance space is the selection of a subset of \textit{impactful} features from the complete feature space. \textit{Feature selection} is an iterative process that starts by extraction of meaningful features from test scenarios, followed by feature pre-processing, removing redundant features, and finally, \textit{feature learning} -- a process that uses machine learning methods to find significant features that clearly distinguish the safe from unsafe scenarios. 

Different features of driving scenarios are measured in different scales and have different ranges. Therefore, we first normalise the features using $z-score$ normalisation. The $z-score$ normalisation scales the features to ensure a mean of zero and a standard deviation of one, and is a known method to re-scale the data and remove the outliers~\cite{birchler2022single, geron2022hands}. The normalised feature set is further refined by correlation analysis. The strongly correlated features indicate \textit{nearly-redundant} information and can cause PCA to overemphasise their contribution. Therefore, such features are removed from the feature set. We further remove the features showing a very weak correlation with the scenario outcome as these features provide little information on why a scenario passes/fails~\cite{hinkle2003applied}. For correlation analysis, we use Spearman Rank Correlation as it is a non-parametric method and does not require assumptions about data distribution or the shape of the relationships, except for the latter being monotonic~\cite{de2016comparing}. The method has been widely used for correlation analysis in previous studies related to AV testing~\cite{aghababaeyan2021black, arrieta2019search, chen2020deep, lu2021search}. 

Once the data is pre-processed, it is time to select a subset of features that have the highest impact on test outcomes. For this, we first identify the clusters of features exhibiting similar properties. To accomplish this, we utilise the k-means clustering algorithm with a dissimilarity measure of $1-|\rho_{i,j}|$, where $\rho_{i,j}$ is the correlation between two features. K-means clustering facilitates a more straightforward interpretation of the clustering outcomes when contrasted with alternative methods~\cite{hauer2020clustering}, and previous research has demonstrated that ISA performs well when employing this clustering technique~\cite{smith2022instance, munoz2017performance, munoz2018instance, mario2020regression}. To determine the optimal number of clusters for k-means, we employ silhouette analysis~\cite{aranganayagi2007clustering}. 

Once the clusters have been established, one feature is selected from each cluster to create a feature set. Assuming that the previous step generated $n$ clusters, each feature set will comprise $n$ features. Each n-dimensional feature set is then projected onto a temporary two-dimensional space using Principal Component Analysis (PCA)~\cite{abdi2010principal}. This process is repeated for all possible combinations of features from all clusters. The coordinates of these temporary two-dimensional spaces serve as input for a set of Random Forest (RF) models, which learn the feature combinations that yield the lowest predictive error when forecasting scenario outcomes. The selected subset of features is used for the generation of the instance space.

\subsubsection{Generation of Instance Space}\label{sec:IS_generation}

Now that the set of most effective features has been identified, we project $nD$ feature space to a $2D$ coordinate system in such a way that the relationship between the features of the test scenarios and their outcomes can easily be identified. An ideal projection is one that creates a linear trend when each feature value and scenario outcome is inspected, i.e., low values of features/scenarios' outcomes at one end of a straight line and high values at the other. Furthermore, the instances that are neighbours in high dimensional feature space should remain as neighbours in the 2D instance space (topological preservation). For this, we use an optimisation method called \textit{Projecting Instances with Linearly Observable Trends} (PILOT )~\cite{munoz2018instance}. The PILOT seeks to fit a linear model for each feature and test outcome, based on the instance location in the $2D$ plane. Mathematically, this involves solving the following optimisation problem:

%
\begin{eqnarray}
	\min					&& \left\| \Ft - \B_{r}\Z \right\|^{2}_{F} + \left\| \Y - \Cbf_{r}\Z \right\|^{2}_{F} 
	\label{eq:optimisation}
	\\
	\text{s.t.}				&& \Z = \A_{r}\Ft \label{eq:score}
	\nonumber
\end{eqnarray}

\noindent where $\Ft$ is the matrix containing the $n$ features of a test scenario, $\Y$ is the column vector containing scenario outcome, $\Z\in\Rbb^{i\times 2}$ is the matrix containing $z_1$ and $z_2$ coordinate values of $i$ scenarios in the $2D$ space, $\A_{r} \in \Rbb^{2 \times n}$ is a matrix that takes the feature values and projects them in $2D$ space, $\B_{r} \in \Rbb^{n \times 2}$ is a matrix that takes the $2D$ coordinates and produces an estimation of the feature values, and $\Cbf_{r}\in\Rbb^{t\times 2}$ is a matrix that takes the $2D$ coordinates and makes an estimation of the technique's performance. In short, Equation~\ref{eq:optimisation} finds the difference between the actual values of the features and performances in a higher dimension and the estimation of these values in $2D$. The lower the difference, the higher the topological preservation. Mathematical proofs and additional technical details of PILOT are available in ~\cite{munoz2018instance}.

\subsection{Predicting Test Scenario Outcome (RQ2)}\label{sec:methodology-rq2}
The features chosen to generate the instance space are selected based on \begin{enumerate*}
    \item how well they explain the test outcome, and
    \item how well they separate pass and fail test scenarios in the $2D$ space. 
\end{enumerate*}. To test the predictive ability of the selected features, we train five Machine Learning (ML) models using the features identified as impactful. The trained models are then used to predict the outcome of previously unobserved driving scenarios without running them in a simulator. The machine learning classifiers used in this study are Random Forest (RF)~\cite{breiman2001random}, K-Nearest Neighbours (KNN)\cite{taunk2019brief}, Decision Tree (DT)~\cite{kotsiantis2013decision}, Multilayer Perceptron (MLP)~\cite{murtagh1991multilayer} and Naive Bayes (NB)~\cite{rish2001empirical}, which have been previously used in similar research works in AV-testing~\cite{kruber2019unsupervised, birchler2022cost, ali2019detection}.

To assess the predictive capabilities of the trained models, it is essential to evaluate them on a fresh set of test instances that have not been encountered during the training phase. For the first case study, to the best of our knowledge, no other existing study has generated an AV testing suite represented in a time series format that allows for the extraction of a similar set of features as the ones utilised in our experiments. The closest available test suite we identified is referenced in ~\cite{lu2023deepscenario}. However, it's worth noting that this dataset lacks certain features that our study has identified as having a significant impact, such as AV\_brake, traffic\_light, and volume of obstacles, among others. Consequently, for this dataset, we divided the test scenarios from the same test suite into two subsets: one for training the machine learning models and the other for testing their predictive performance. The models were trained on 80\% of the data, while their predictive performance was assessed using the remaining 20\% of the data, which had not been used during the training phase. 

For the second case study for our experiments, we generated two different test suites using two test generation strategies, frenetic~\cite{frenetic} and AmbieGen~\cite{humeniuk2022ambiegen}, presented in the SBST tool competition~\cite{gambi2022sbst}. The details of these test generation techniques and the test suites generated by them are given in Section~\ref{sec:dataset}. We train the models on the Frenetic test suite using the features selected by ISA. However, to test the model's generalisation ability on uncovered regions and to prove the effectiveness
of the trained model using selected features, the models are evaluated using a test suite generated by AmbieGen. 

We use Python's scikit-learn~\cite{kramer2016scikit} library for the implementation of the machine learning models. The performance is measured in terms of precision, recall, and f1-score. For the comparison, we use random feature selection as a baseline to gauge the performance of our proposed technique.

\subsection{Coverage of Scenario Space Using Boundary Analysis (RQ3)}
\label{sec:boundary}

The instance space represents the test scenarios as points in a $2D$ space. The convex hull of these points represents the area in which test instances exist in the test suite under study. However, the test suite doesn't necessarily include all the test instances required to explore the feature space completely. Using ISA, we devise a way to compute a mathematical boundary that encloses all the test scenarios which are empirically possible to generate, though, may be missing from the current instance space/test suite. 

We compute the boundary as follows: Let $\mathbb{R}^{n\times n}$ be the correlation matrix of $n$ features, characterising a test scenario. We define two vectors  $\mathbf{f}_U = \begin{bmatrix} f_{U_1} \cdots f_{U_n} \end{bmatrix}^T$ and $\mathbf{f}_L = \begin{bmatrix}f_{L_1} \cdots f_{L_n}\end{bmatrix}^T$
containing the upper and lower bounds of feature values. From these vectors, we define a vertex vector containing a combination of values from $f_U$ and $f_L$ such that only the upper or lower bound of a feature is included. For instance, $\mathbf{v}_1 = \begin{bmatrix}f_{U_1} f_{L_2} \cdots f_{L_n}\end{bmatrix}^T$ represents a vertex vector containing the maximum value of feature 1 and minimum values of all the other features. We define a matrix $\mathbf{V} = \begin{bmatrix}\mathbf{v}_1 \cdots \mathbf{v}_q\end{bmatrix} \in \mathbb{R}^{n\times q}, q = 2^n$ containing all possible vertices created by the feature combinations. The vertices in metric $\mathbf{V}$, connected by edges, define a hyper-cube that encloses all the instances in the instance space. 

Some of the vectors in $\mathbf{V}$ represent feature combinations that are unlikely to coexist. For example, if features 1 and 2 are strongly positively correlated, it is unlikely to find instances that have a high value of feature 1 and a low value of feature 2. Therefore, a vertex vector $\mathbf{v} = \begin{bmatrix}f_{U_1} f_{L_2} \cdots f_{L_n}\end{bmatrix}^T$ would be unlikely to be near any true instance. Similarly, a vertex vector cannot simultaneously contain $\{f_{U_1}, f_{U_2} \}$ or $\{f_{L_1}, f_{L_2} \}$ if feature 1 and feature 2 are strongly negatively correlated. All such unlikely vertex vectors are eliminated from $\mathbf{V}$. The edges connecting the remaining vertex vectors are then projected into $2D$ instance space using PILOT~\cite{munoz2018instance}, whose convex hull now represents the mathematical boundary of the expanded instance space. 

We propose a new coverage measure, called Instance Space Coverage ($\texttt{Coverage}_{IS}$), that measures the percentage of the area of the \textit{possible scenario space} ($\texttt{area}_{bound}$) -- defined by the mathematical boundary -- that is covered by the generated instance space ($\texttt{area}_{IS}$). 

\begin{equation}
\label{eq:coverage}
\texttt{Coverage}_{IS} (\%) = \frac{\texttt{area}_{IS}}{\texttt{area}_{bound}} \, 100 
\end{equation}

For calculating the $area_{IS}$, we focus exclusively on the high-density regions within the instance space to ensure that the calculation remains unaffected by outliers. To achieve this, we employ the DBSCAN clustering algorithm~\cite{schubert2017dbscan}, which defines a cluster as a densely populated area of data points. DBSCAN requires two input parameters, denoted as $\left\{k,\varepsilon\right\}$, where $k$ signifies the minimum number of instances necessary to establish a dense region, and $\varepsilon$ denotes the maximum radius within which an instance is considered a neighbour. The Euclidean distance metric is used for distance measurement. The values for $k$ and $\varepsilon$ are determined automatically following the recommendations in~\cite{daszykowski2001looking}, employing Equations~\ref{eq_DBSCAN_params1} and~\ref{eq_DBSCAN_params2}.

\begin{equation}
k \leftarrow \max\left(\min\left(\left\lceil r/20\right\rceil,50\right),3\right)
\label{eq_DBSCAN_params1}
\end{equation}

\begin{equation}
\varepsilon \leftarrow \frac{k\Gamma\left(2\right)}{\sqrt{r\pi}}
\left(\text{range}\left(z_{1}\right)\times\text{range}\left(z_{2}\right)\right)
\label{eq_DBSCAN_params2}
\end{equation}

Where $r$ represents the number of scenarios, $\Gamma\left(\cdot\right)$ is the Gamma function, and ${z_1, z_2}$ are the coordinates within the two-dimensional instance space.

To delineate the footprint of each cluster, we employ an $\alpha$-shape, a concept in computational geometry that extends the notion of a convex hull~\cite{edelsbrunner1983shape}. It essentially defines a polygon that closely envelops all the data points within a cluster. An $\alpha$-shape is generated for each cluster, and these shapes are collectively bounded together as a MATLAB polygon structure~\cite{polygons}. Finally, we compute the area of the polygon using MATLAB's polyarea function~\cite{Polyarea}.

For $area_{bound}$, we compute the polygon using boundary instances as vertices and calculate the area occupied by the polygon.

\section{Experimental Results}\label{sec:results}
This section reports the insights
we have gained through the visualisations of the distributions of features in the instance space generated by the test suites used in our study, and seek the answers to the research questions introduced in Section~\ref{sec:introduction}. 

\begin{figure}[!t]
    \centering
    \includegraphics[width=0.6\linewidth]{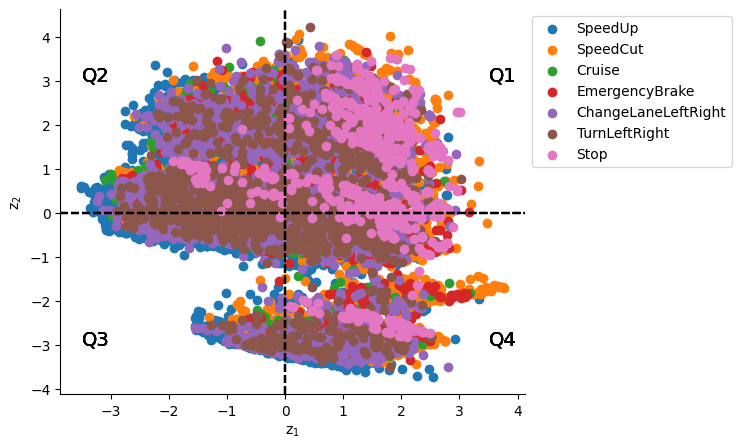}
    \caption{Instance Space for test scenarios labelled by the operation performed by AV}
    \label{fig:sources}
\end{figure}

\begin{figure*}[!ht]
    \subfloat[][AV\_brake]{\includegraphics[width=0.29\linewidth]{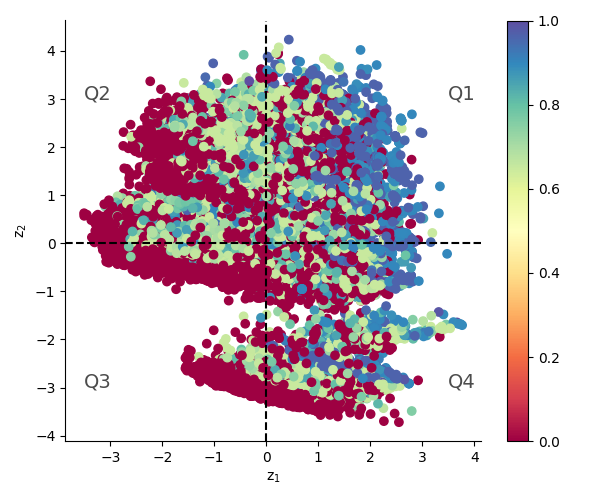}\label{fig:feature_ego_brake}}%
	\subfloat[AV\_speed]{\includegraphics[width=0.29\linewidth]{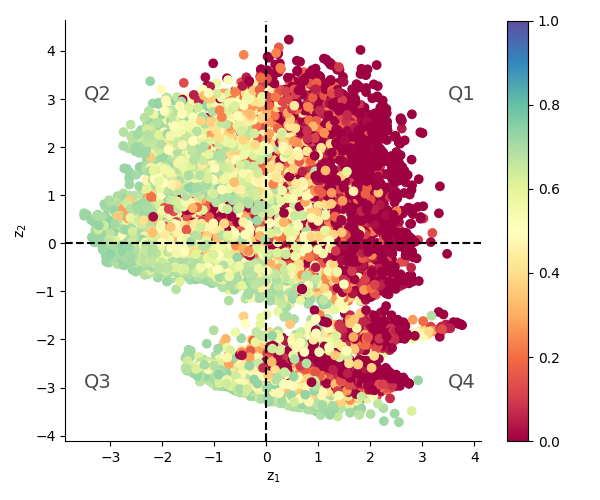}\label{fig:feature_ego_speed}}%
	\subfloat[is\_ped]{\includegraphics[width=0.29\linewidth]{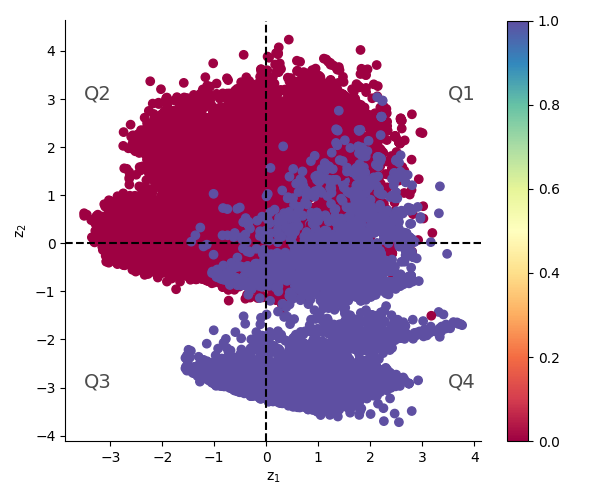}\label{fig:feature_is_pedestrian_scenario}}\\%
	
	\subfloat[num\_NPCs]{\includegraphics[width=0.29\linewidth]{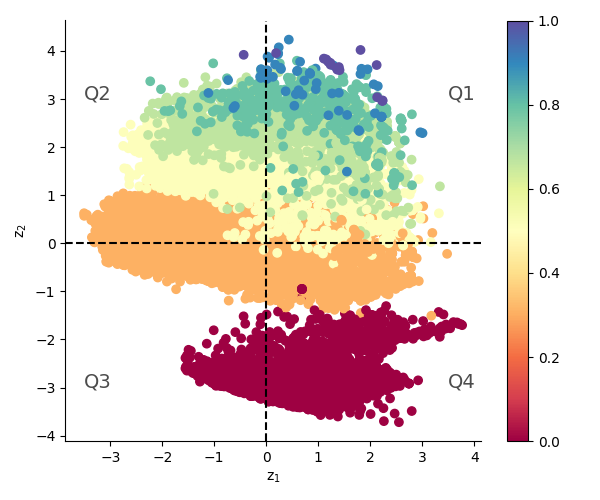}\label{fig:feature_total_NPCs}}%
	\subfloat[num\_obs]{\includegraphics[width=0.29\linewidth]{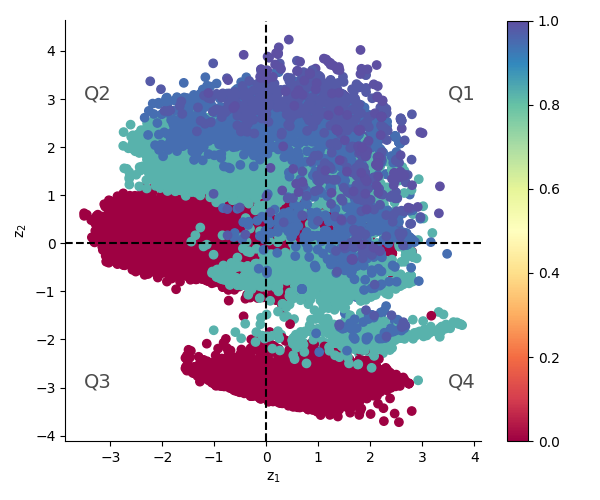}\label{fig:feature_total_obstacles}}%
	\subfloat[traffic\_light]{\includegraphics[width=0.29\linewidth]{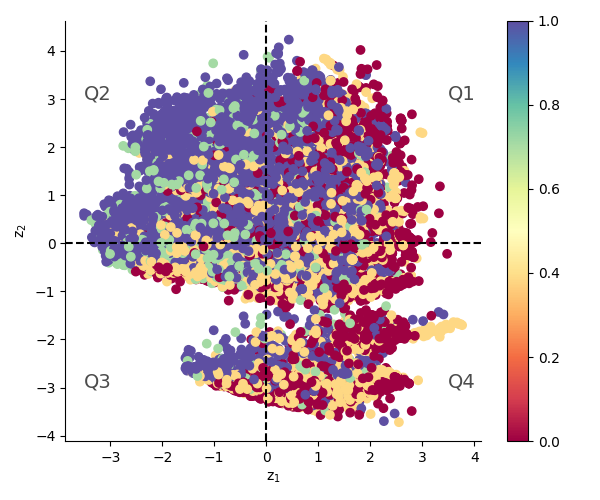}\label{fig:ftraffic_light}}\\%
	
	\subfloat[dist\_maxSpeedObs]{\includegraphics[width=0.29\linewidth]{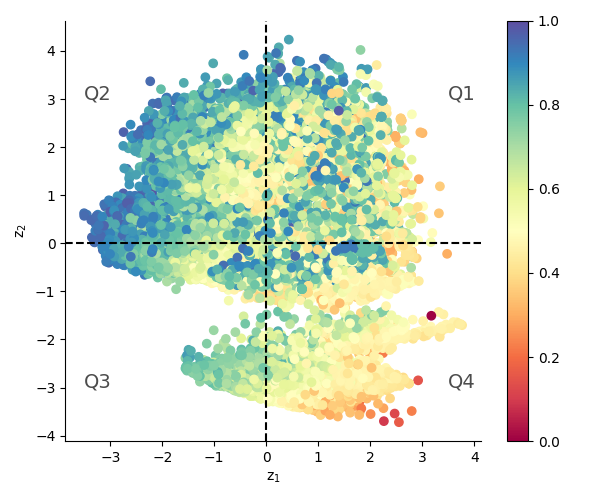}\label{fig:feature_dist_maxSpeedObstacle}}%
	\subfloat[min\_obsDist]{\includegraphics[width=0.29\linewidth]{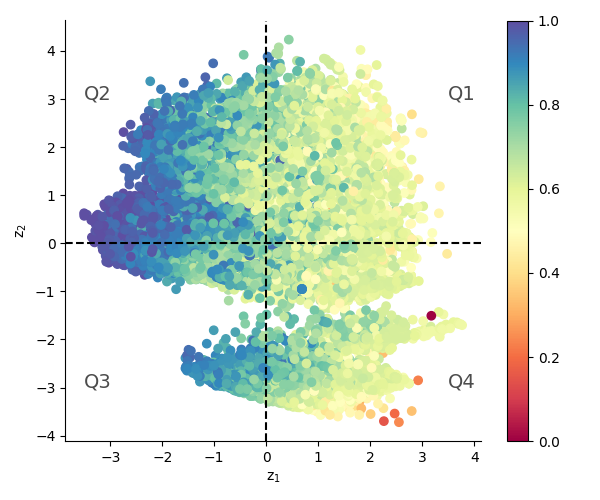}\label{fig:feature_min_obstaclesDistances}}%
	\subfloat[speed\_minDistObs]{\includegraphics[width=0.29\linewidth]{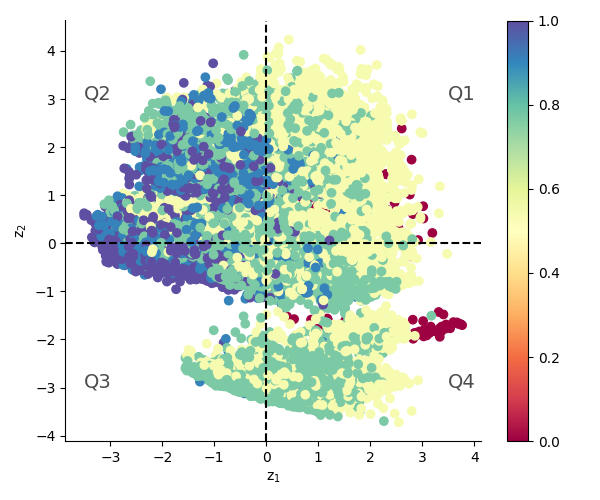}\label{fig:feature_speed_minDistObstacle}}\\%
	
	\subfloat[vol\_minDistObs]{\includegraphics[width=0.29\linewidth]{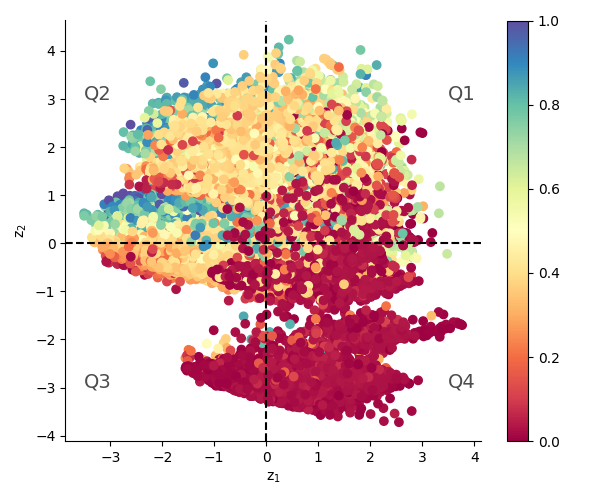}\label{fig:feature_vol_minDistObstacle}}%
 \subfloat[scenario outcome]
 {\includegraphics[width=0.29\linewidth]
 {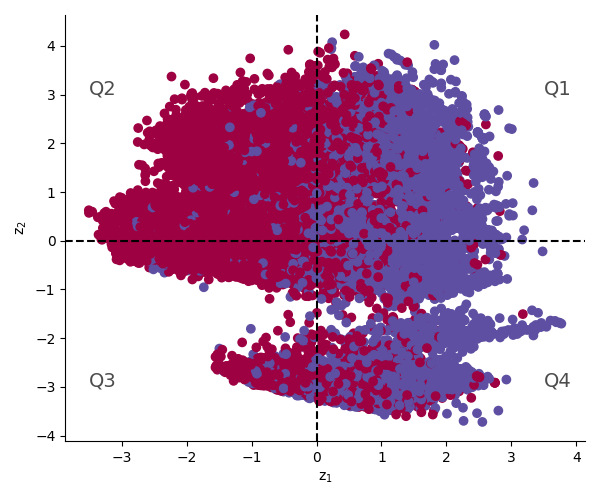}\label{fig:scenario_performance_ds1}}
	\caption{(a -- j) The distribution of the features in the instance space, from lower values in red to higher values in blue. (k) Red represents safe while blue represents unsafe scenarios}
     \label{fig:feat_dist}
\end{figure*}

\subsection{Key features of effective test scenarios (RQ1)}
\label{sec:RQ1}

\subsubsection{Dataset 1}
ISA selected ten features that have an impact on the effectiveness of test scenarios. These are  \texttt{AV\_brake} and \texttt{AV\_speed}, which represent the initial speed and brake applied by the AV; \texttt{num\_NPCs}, \texttt{num\_obs} and \texttt{traffic\_light}, and \texttt{Is\_ped} define the background of the scenario and are categorised under the static features. The remaining four features report the speed, distance, and volume of the obstacles and other road users which are part of the scenario. The selected features are then projected to an $2D$ instance space (Figure~\ref{fig:sources}), and the projection matrix defined by the linear transformation is shown in Equation~\ref{eq:pcs}. The projection matrix shows the contribution of each feature to the $z_1$ and $z_2$ axis. \texttt{Is\_ped} has the highest contribution to $z_1$ axis, while \texttt{num\_NPCs} contributed the most to $z_2$ axis. 

The distribution of the selected features in the instance space is shown in figure~\ref{fig:feat_dist} (a -- j). The features show a gradual increasing/decreasing trend from one edge of the space to the other. For $traffic\_light$, the minimum value shows the absence of any traffic signal on the road, followed by green, orange, and red (maximum value).

\begin{equation}
\label{eq:pcs}
    \begin{bmatrix}
    	z_1 \\
            z_2
    \end{bmatrix} = 
     \begin{bmatrix}
    	{0.18} &   {0.11}   \\
            {-0.46} &   {-0.12} \\
            {-0.21} &   {0.13}  \\
            {0.03}  &   {0.64}  \\
            {0.26}  &   {-0.46} \\
            {0.14}  &   {0.44}  \\
            {-0.51} &   {0.04}  \\
            {-0.32} &   {-0.06} \\
            {-0.13} &   {0.37}  \\
            {-0.26} &   {0.22}
    
    \end{bmatrix}^{T}
     \begin{bmatrix}
    	{\text{AV\_brake}}  \\
    	{\text{AV\_speed}} \\
    	{\text{traffic\_light}}\\
    	{\text{num\_NPCs}}\\
    	{\text{Is\_ped}}\\
    	{\text{num\_obs}}\\
    	{\text{min\_obsDist}}\\
    	{\text{speed\_minDistObs}}\\
    	{\text{vol\_minDistObs}}\\
            {\text{dist\_maxSpeedObs}}
    \end{bmatrix}
\end{equation}

The features \texttt{AV\_speed}, \texttt{traffic\_light}, \texttt{min\_obsDist} and \texttt{speed\_minDistObs} show a decreasing trend from left to right of the instance space, and these features except \texttt{traffic\_light} divide the space horizontally with their low and high values. Contrary to these features, the values of \texttt{AV\_brake} increase from left to right. The other four features show a gradual change in their values from top to bottom. For \texttt{num\_NPCs}, \texttt{num\_obs} and \texttt{vol\_minDistObs}, the lowest values are at the bottom, whereas for \texttt{is\_ped} the lowest values are at the top. The distribution trend of the feature \texttt{dist\_maxSpeedObs} isn't very well-defined and thus can not be used to deduce any interesting results.  

Figure~\ref{fig:scenario_performance_ds1} shows the distribution of test scenarios according to their outcome. The scenarios that lead to a collision or break the safety distance are represented by blue dots, while red dots represent safe scenarios, where the AV travelled from the source to the destination without colliding or breaking the minimum safety requirement. 
It can be seen that the critical scenarios are majorly located in Q1 and Q4, we, therefore, are interested in identifying the defining features of the scenarios located in these quadrants. By superimposing the feature distributions over scenario outcomes, we can identify the features that impact the effectiveness of these scenarios. 

The distribution of features \texttt{AV\_brake}, \texttt{AV\_speed}, \texttt{traffic\_light}, \texttt{min\_obsDist}, and \texttt{speed\_minDistObs} show a similar trend as the distribution of scenario outcome (increasing/decreasing trend from left to right), thus showing a strong impact on the outcome of test scenarios. In the critical region of the space, the value of the brake applied by the AV is high (green and blue dots), which means the AV exhibited sudden and hard braking. Although there are exceptions, high values are dominant (Figure~\ref{fig:feature_ego_brake}). Contrary to the brake, a lower initial speed of the AV leads to more safety-critical scenarios (Figure~\ref{fig:feature_ego_speed}). For the feature \texttt{traffic\_light}, either it is green or there is no signal at all, whereas, Low to medium values of \texttt{min\_obsDist} and \texttt{speed\_minDistObs} are prominent in Q1 and Q4 (Figure~\ref{fig:feature_min_obstaclesDistances} --~\ref{fig:feature_speed_minDistObstacle}). Furthermore, it can be seen that the number of obstacles is high in the safety-critical region of the space (Figure~\ref{fig:feature_total_obstacles}), showing that scenarios simulating busy roads are more effective. In summary, the most contributing features explaining test outcomes are \texttt{AV\_brake}, \texttt{AV\_speed}, \texttt{num\_obs}, \texttt{traffic\_light}, \texttt{min\_obsDist}, and \texttt{speed\_minDistObs}. 

Figure~\ref{fig:feat_dist} shows the impact of the features on the effectiveness of test scenarios individually. However, as AV operates in an environment having many of these features together, inspecting the combined impact of features on the scenario's performance would explain the impact more meaningfully. We list some examples of the effective combinations of features that appear to contribute together to defining the safety-criticality of the scenarios:

\begin{itemize}
\item The impact of hard braking seems to correlate with the traffic signal. When there is no signal on the road, or the traffic light is green, sudden braking leads to collision (Q1, Q4 of Figure~\ref{fig:scenario_performance_ds1}). On the other hand, in the scenarios where the traffic signal is red (blue clusters in Q2 of Figure~\ref{fig:ftraffic_light}), hard braking scenarios appear safe (green clusters in Q2 of Figure~\ref{fig:feature_ego_brake}). 

The scenario that explains this observation is that when the traffic signal is green or there is no road signal, other vehicles are driving at the road speed. The sudden brake applied by AV could lead to a rear-end collision with other vehicles. On the other hand, if the signal is red, other vehicles are already stopped or driving at a very low speed (preparing to stop), and thus the chances of a rear-end collision with an AV are low. 

\item The initial speed with which AV starts its journey on the road (\texttt{AV\_speed}) and traffic signal also seem to have a combined impact on the test scenario's outcome. The regions of the instance space where the initial speed of AV is slow, and the traffic light is green (or no traffic light), show a higher number of safety-critical scenarios. 

We observe this combined impact in the scenarios where AV is parked parallel on the side of the road. If other vehicles are driving at the road's designated speed (no or green signal), it is important for AV to catch the speed of other vehicles as soon as it gets out of the parking bay to avoid collision with the vehicles coming from behind. However, in the presence of a red or orange signal, either other vehicles are already stopped or slowed down, thus reducing the chances of collision with the AV.

\item The feature \texttt{speed\_minDistObs} shows the speed of an obstacle that is at the minimum distance from the AV. Intuitively, higher speed should be more accident-prone. However, the highest values of this feature belong to the instances in Q3 of Figure~\ref{fig:feature_speed_minDistObstacle}, which contains the majority of safe instances. This behaviour can be explained by combining the impact of this feature with \texttt{vol\_minDistObs}, which shows the volume of these high-speed obstacles. By comparing~\ref{fig:feature_speed_minDistObstacle} and~\ref{fig:feature_vol_minDistObstacle}, it can be seen that the high volume of these high-speed obstacles prevents them from colliding with AV. From Figure~\ref{fig:sources} it can be seen that most of the scenarios in this region come under \texttt{Lane Change} and \texttt{Speed Cut} operations of AV.  Therefore, if the volume of the closest vehicles is high, the sensors easily detect the obstacles even in the blind spot while changing lanes, thus reducing the chances of collision. 
\end{itemize}

From these results, we infer that although features impact the outcome of a test scenario individually, it is the combination of critical features that fully defines their safety-criticality.  

\subsubsection{Dataset 2}: This test suite has been specifically generated to evaluate the lane-keeping functionality of an autonomous vehicle (AV). It comprises a smaller set of 15 features in comparison to test suite 1. These 15 features are derived from five primary features, which include the angle of the turns, the radius of the turns, the number of turns, the number of straight segments, and the total length of the road. The remaining features are computed statistically from these five base features, encompassing calculations for the minimum value, maximum value, mean value, median value, and standard deviation.

ISA identified three out of these features as having the most significant impact on scenario outcomes. These key features are \textit{median\_angle}, \textit{num\_r\_turns}, and \textit{road\_distance}. These features have been projected into a $2D$ space using the methodology described in Section~\ref{sec:IS_generation}. The projection matrix, as presented in Equation~\ref{eq:pcs2}, illustrates the contribution of each feature to the $z\_1$ and $z\_2$ axes. Notably, \textit{road\_distance} plays a predominant role in the $z\_1$, while \textit{num\_r\_turns} exerts the greatest influence on the $z\_2$ axis.

\begin{equation}
\label{eq:pcs2}
    \begin{bmatrix}
    	z_1 \\
            z_2
    \end{bmatrix} = 
     \begin{bmatrix}
     {0.3726}  &   {-0.4046}\\
     {-0.2019}    &   {0.7063}\\
     {0.614}   &   {0.5738}

    \end{bmatrix}^{T}    
     \begin{bmatrix}
    	{\text{median\_angle}}  \\
    	{\text{num\_r\_turns}}\\
    	{\text{road\_distance}}
    \end{bmatrix}
\end{equation}
\begin{figure}[H]
    \subfloat[][median\_angle]{\includegraphics[width=0.32\linewidth]{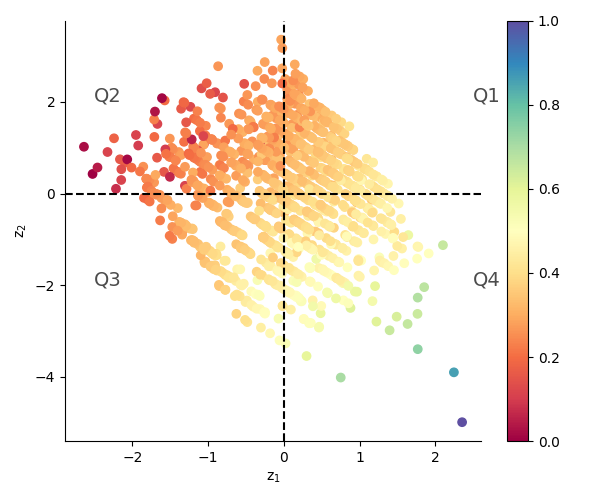}\label{fig:feature_median_angle}}%
	\subfloat[num\_r\_turns]{\includegraphics[width=0.32\linewidth]{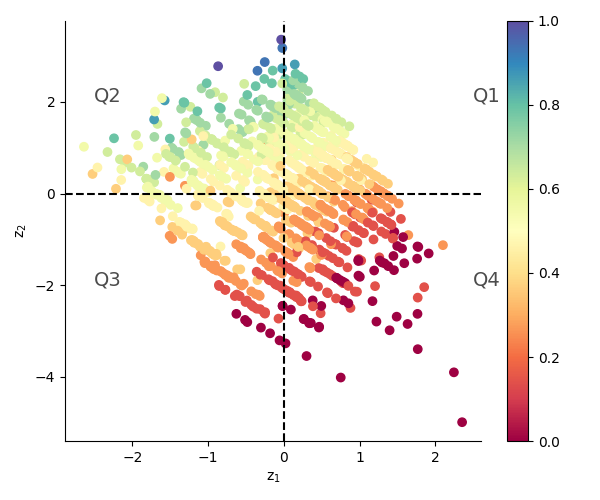}\label{fig:feature_num_r_turns}}%
	\subfloat[road\_distance]{\includegraphics[width=0.32\linewidth]
 {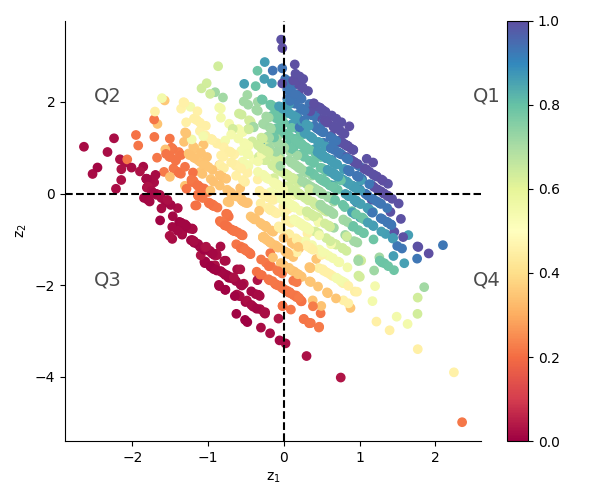}\label{fig:feature_road_distance}}\\%
 \subfloat[scenario\_outcome]{\includegraphics[width=0.32\linewidth]
 {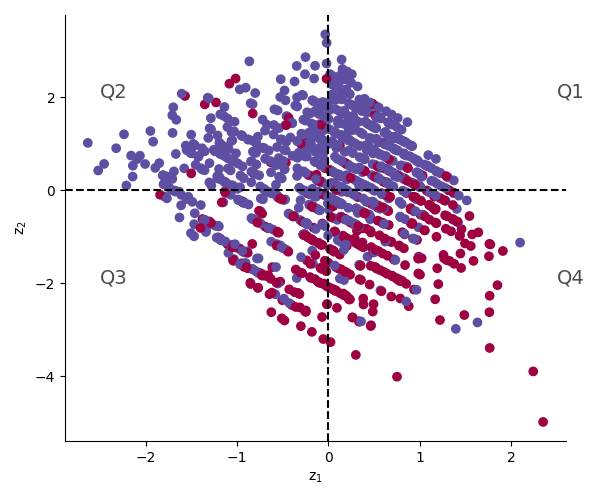}\label{fig:feature_scenario_outcome2}}%
\caption{(a -- c) The distribution of the features in the instance space, from lower values in red to higher values in blue. (d) Red represents safe while blue represents unsafe scenarios}
     \label{fig:feat_dist}
\end{figure}

Figures~\ref{fig:feature_median_angle} -- \ref{fig:feature_road_distance} display the distribution trends of these features within the instance space. The gradual transition in colour, from red to blue, signifies the progressive change in feature values across the space. Additionally, Figure~\ref{fig:feature_scenario_outcome2} illustrates the distribution of scenario outcomes. Instances where the AV deviates from the centre of the lane, denoting out of bound episodes, are represented by blue dots, while red dots signify safe driving scenarios.

To better understand the impact of each feature on the outcome, one can overlay the feature distribution with the scenario outcome distribution. Notably, the values of \textit{num\_r\_turns} exhibit an ascending pattern from the bottom to the top diagonally, aligning with the scenario outcome distribution, which indicates a higher incidence of failure along the same diagonal. Consequently, roads featuring a higher number of right turns pose increased challenges for the AV in maintaining lane boundaries. Similarly, sharp turns, characterised by medium-to-low values of \textit{median\_angle}, are associated with failed test scenarios.

It's essential to clarify that all features undergo normalisation between 0 and 1 before being projected into the instance space. Therefore, a value of 0 for \textit{median\_angle} does not indicate a 0-degree turn angle, but instead signifies the minimum value of the turn angle within the test suite.

The feature \textit{road\_distance} exhibits a bottom-to-top progression that corresponds to the scenario outcome distribution. However, unlike the first two features, this mapping is less distinct. The projection method, explained in Section~\ref{sec:IS_generation}, optimises based on two primary goals: \begin{enumerate*}
    \item topological preservation; and
    \item a clear separation of safe and unsafe instances.  
    \end{enumerate*}
As a result, a feature that may not exhibit a clear one-to-one mapping with the test outcome can still be selected based on the optimisation criteria of PILOT.

\subsection{Predicting test scenario outcomes (RQ2)}

\begin{table*}[!ht]
    \centering
    \caption{The performance of the Random Forest Classifier (RFC), Decision Tree (DT), K-Nearest Neighbours (KNN), Multi-Layer Perceptron (MLP) and Naive Bayes (NB) classifiers in terms of precision (P), recall (R) and F1 score (F1). Rows labelled as \textit{ISA} and \textit{Random} report the performance of classifiers using features selected through ISA or picked randomly, respectively.}
    \label{tab:classifiers_performance}
    \begin{tabular}{ll|lll|rrr}
\toprule

                    &       &   \multicolumn{3}{c}{Dataset1}    &   \multicolumn{3}{c}{Dataset2}\\
                    &       &   ISA  &   Random     & p-value   & ISA       & Random    &   p-value\\\hline

\multirow{3}{*}{RF}	&	  P	&	\textbf{0.876}	&	0.787	&	0.002	&	\textbf{0.796}	&	0.692	&	0.002	\\
                    &	R	&	\textbf{0.868}	&	0.780	&	0.002	&	0.757	&	0.695	&	0.084	\\
                    &	F1	&	\textbf{0.870}	&	0.781	&	0.002	&	\textbf{0.774}	&	0.675	&	0.002	\\\hline

\multirow{3}{*}{DT}	&	P	&	\textbf{0.847}	&	0.780	&	0.002	&	\textbf{0.788}	&	0.695	&	0.002	\\
                    &	R	&	\textbf{0.839}	&	0.768	&	0.002	&	\textbf{0.825}	&	0.692	&	0.020	\\
                    &	F1	&	\textbf{0.841}	&	0.769	&	0.002	&	\textbf{0.804}	&	0.671	&	0.002	\\\hline

\multirow{3}{*}{KNN} &	P	&	\textbf{0.855}	&	0.780	&	0.004	&	\textbf{0.806}	&	0.664	&	0.002	\\
                    &	R	&	\textbf{0.851}	&	0.770	&	0.004	&	\textbf{0.818}	&	0.670	&	0.002	\\
                    &	F1	&	\textbf{0.852}	&	0.772	&	0.004	&	\textbf{0.812}	&	0.655	&	0.002	\\\hline

\multirow{3}{*}{MLP} &	P	&	\textbf{0.864}	&	0.753	&	0.002	&	\textbf{0.787}	&	0.702	&	0.002	\\
                    &	R	&	\textbf{0.865}	&	0.753	&	0.002	&	\textbf{0.795}	&	0.679	&	0.020	\\
                    &	F1	&	\textbf{0.865}	&	0.751	&	0.002	&	\textbf{0.791}	&	0.675	&	0.002	\\\hline

\multirow{3}{*}{NB} &	P	&	\textbf{0.806}	&	0.730	&	0.002	&	\textbf{0.807}	&	0.624	&	0.002	\\
                    &	R	&	\textbf{0.796}	&	0.683	&	0.002	&	\textbf{0.832}	&	0.646	&	0.002	\\
                    &	F1	&	0.799	&	0.676	&	0.160	&	\textbf{0.819}	&	0.621	&	0.002	\\\hline

    \end{tabular}

\end{table*}

To evaluate the adequacy of the features selected for the generation of the instance space, we test their performance in terms of predicting the scenario outcome without them being executed in the simulator. For this, we train five machine learning classifiers using these features and use the trained models to classify the unknown scenarios either as safe or unsafe. For comparison, we also train these classifiers with randomly selected features and present their predictive performance. Table~\ref{tab:classifiers_performance} reports the performance in terms of precision, recall and F1 score. To find the statistical difference in the performance by using different sets of features, we use Wilcoxon Signed-Rank test \textit{p}-values. The \textit{p}-value $\leq 0.05$ indicates a significant difference between the quality of the solutions provided by the different methods, insignificant otherwise. 

As discussed in Section~\ref{sec:methodology-rq2}, for Dataset 1, the model training is performed on 80\% of the data, while the rest 20\% is used for testing. However, for Dataset 2, training and testing are performed on two different datasets. Models are trained on the test scenarios generated using \textit{Frenetic} test generation technique~\cite{frenetic}, while tested on a test suite generated by \textit{AmbieGen} test generator~\cite{humeniuk2022ambiegen}. Evaluating the classification models on a different test suite, which is not part of the instance space, gives us an opportunity to test the generalizability of the proposed method.

It can be seen that for both of the datasets, the models trained with the features selected by our methodology perform statistically better than the features chosen randomly. These features, therefore, can be used in test case selection and prioritisation techniques, where critical scenarios can be identified based on \textit{impactful features}, without executing the scenarios in a simulator. Furthermore, though the randomly selected features do not perform as well as the features selected by our proposed methodology, their performance is not very low. In the majority of cases, the precision, recall and F1 score of the classifiers trained with random features is higher than 70\% for dataset1, and 60\% for dataset 2. This gives us further confidence in the suitability and relevance of the set of features we extracted for our study. 


\subsection{Instance Space Coverage (RQ3)}

\begin{figure*}[!ht]
    \subfloat[]{\includegraphics[width=0.45\linewidth]{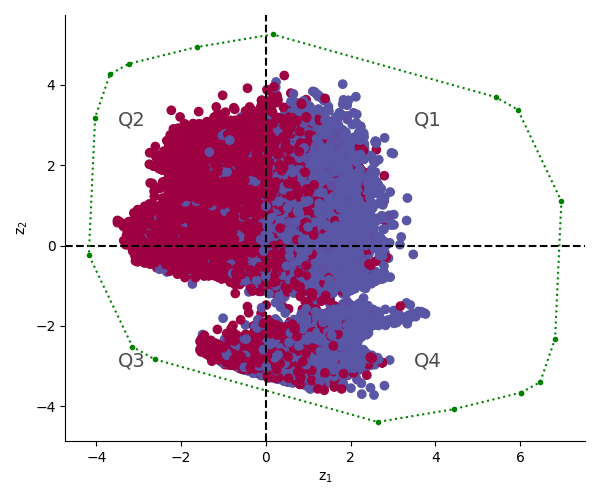}\label{fig:boundary_scenario_performance_ds1}}%
    \subfloat[]{\includegraphics[width=0.45\linewidth]{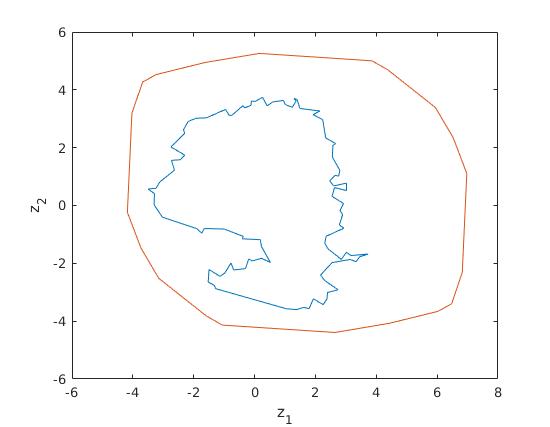}\label{fig:polygon_area_ds1}}%

	\caption{Dataset1: (a) Boundaries surrounding all the possible test scenarios. Red represents safe, while blue represents safety-critical scenarios. (b) Polygons generated by the expanded boundary (red) and the instance space (blue)}
     \label{fig:boundaries_ds1}
\end{figure*}

\begin{figure*}[!ht]
    \subfloat[]{\includegraphics[width=0.45\linewidth]{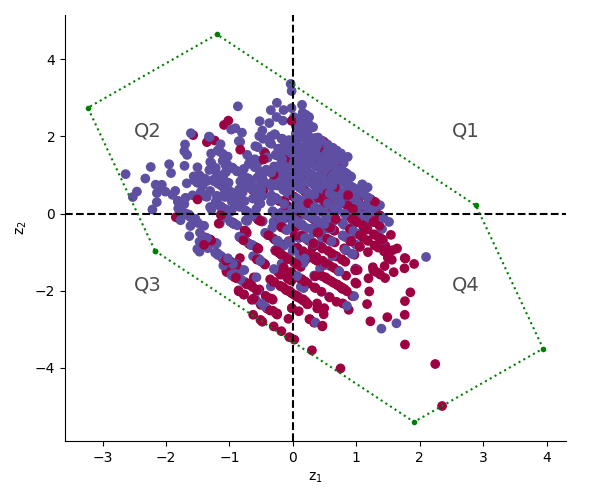}\label{fig:boundary_scenario_performance_ds1}}%
    \subfloat[]{\includegraphics[width=0.45\linewidth]{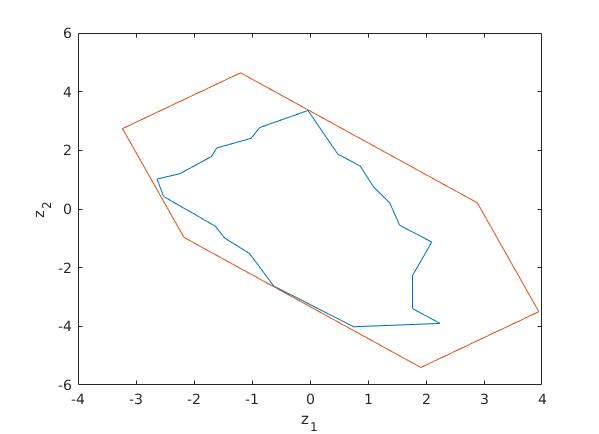}\label{fig:polygon_area_ds2}}%
  
	\caption{Dataset 2: (a) Boundaries surrounding all the possible test scenarios. Red represents safe, while blue represents safety-critical scenarios. (b) Polygons generated by the expanded boundary (red) and the instance space (blue)}
     \label{fig:boundaries_ds2}
\end{figure*}

The coverage metric introduced in Section~\ref{sec:boundary} provides an objective measure of how extensively the scenario space has been explored relative to the expanded boundary computed mathematically. Figures~\ref{fig:boundaries_ds1} and~\ref{fig:boundaries_ds2} show the boundary created based on the maximum and minimum values of the features deemed as impactful by ISA for dataset 1 and dataset 2 respectively. The empty regions enclosed by the boundary represent test scenarios that are empirically possible to generate, however, they are not part of the current test suite.

Figures~\ref{fig:polygon_area_ds1} and~\ref{fig:polygon_area_ds2} show the polygons generated by the expanded boundary (red) and the instance space (blue) for Dataset 1 and 2 respectively. For Dataset 1, the areas computed for these polygons are: $\texttt{area}_{IS}  = 29.84 $ and $\texttt{area}_{bound}  = 91.28 $ respectively. Using equation~\ref{eq:coverage}, the coverage of this test suite is computed as 32.69\%, indicating that 67.33\% of the possible feature space is still untested. For Dataset 2, $\texttt{area}_{IS}  = 18.40 $ and $\texttt{area}_{bound}  = 36.79 $, resulting in test coverage of 50.01\%. 

Striving for 100\% coverage by generating driving scenarios to test every possible driving situation is impractical, as simulating test scenarios is a resource and time-intensive activity. Therefore, more effort should be put into testing the error-prone regions of the testing space. As the instance space is created in such a way that safe scenarios lie at one end of the space and unsafe at the other, it is possible to find the regions of the feature space where the probability of failure is high. For Dataset 1,  based on the distribution trend for scenario outcome, the probability of failure is higher in  Q1 and Q4. For Dataset 2, as the distribution pattern of failure increases from bottom-right to top-left (diagonally), Q2 seems to have a high probability of failed scenarios. Adding new scenarios to these regions will improve the quality of testing by adding more challenging conditions for the system.

Generation of the test scenarios having feature values that will project to a particular point in the instance space is a challenging task. One possible way of doing this is by setting the target points in the instance space, and evolving existing instances, using Genetic Algorithms, to adapt their data so that they occupy the target locations in the instance space~\cite{munoz2020generating}. However, our focus in this study is to assess how well the scenario space has been explored, which is computed using the coverage of the instance space. The generation of the missing test scenarios using controllable test properties is beyond the scope of the current study, and we intend to pursue this as future research.  

\section{Related Studies}\label{sec:related-work}
In the realm of AV testing, the representation of scenarios poses a challenge due to the intricate nature of the search space and the presence of an infinitely large number of parameter combinations, as highlighted in the survey by Ding et al.~\cite{ding2022survey}. This section aims to explore the existing studies in this field, shedding light on prior research efforts. Furthermore, it elucidates how the proposed technique differs from the prior work, emphasising the novel contributions and advancements it brings to the domain.

\subsection{Safety Critical Scenario Generation}
A study conducted through surveys and semi-structured interviews with developers and testers working in the AV industry identifies that generating possible corner cases and unexpected driving scenarios is a crucial requirement in AV testing~\cite{lou2022testing}. Several techniques have been proposed to generate safety-critical test scenarios for AVs~\cite{chandrasekaran2021combinatorial, gambi2019generating, han2020metamorphic, kluck2018using, li2020ontology, tao2019industrial, tian2018deeptest, zhang2018deeproad, zhou2020deepbillboard, tian2022generating, abdessalem2018testing, ben2016testing, calo2020generating}. Chandrasekaran et al. propose a combinatorial testing approach to generate images to test DNN models used in AVs ~\cite{chandrasekaran2021combinatorial}. AC3R uses natural language processing and domain-specific ontology to extract car crash driving scenarios from police reports~\cite {gambi2019generating}. Kluck et al. use domain ontologies providing the environmental models, and combinatorial testing for obtaining critical scenarios~\cite{kluck2018using}. In~\cite{zhang2018deeproad, tian2018deeptest,han2020metamorphic}, metamorphic relations are used to test the behaviour of an ADS. Zhou et al. generate safety-critical scenarios by adding adversarial perturbations on real-world billboards~\cite{zhou2020deepbillboard}. Tian et al. mine behaviour patterns that lead to collision from naturalistic driving datasets. Utilising these patterns, they first create abstract scenarios by combining various behavioural patterns and then generate concrete scenarios by calculating participant trajectories that align with the selected patterns~\cite{tian2022generating}. Stocco et al. present an architecture in their papers to identify safety-critical misbehaviour, such as collisions and out of bound occurrences, in autonomous driving systems~\cite{stocco2020misbehaviour, stocco2022confidence}. The proposed framework involves using a variational autoencoder (VAE) to reconstruct a sequence of prior input images of a current scene and to determine the corresponding reconstruction errors. The model is trained on normal data and learns a probability distribution of the observed reconstruction errors using maximum likelihood estimation. This distribution is then utilised to establish a threshold value for differentiating between anomalous and normal behaviour while regulating the detection's false positive rate. An SDC, equipped with the trained model, raises a misbehaviour flag when it encounters a scenario that yields a high reconstruction error compared to the threshold computed during training.

Search-based techniques are well suited for system testing of complicated problems like AVs~\cite{zeller2017search}. SAMOTA~\cite{haq2022efficient} combines surrogate-assisted optimisation and many-objective search to generate test scenarios effectively and efficiently. AsFault~\cite{gambi2019automatically} uses a genetic algorithm to search for the most properties of a road that lead AV to move away from the centre of the lane. In~\cite{abdessalem2018testing}, Abdessalem et al. use a multi-objective evolutionary algorithm to find the system-level failures of AVs, caused by feature interaction. In another study~\cite{ben2016testing}, they use surrogate models with multi-objective search to test the pedestrian detection system of an ADS. AV-Fuzzer~\cite{av-fuzzer}, MOSAT~\cite{tian2022mosat} and ATLAS~\cite{Tang2021} use Genetic algorithms to manipulate the manoeuvres of other vehicles on the road (Non-playable Characters -- NPCs) to make the driving scenarios more challenging for AVs. AutoFuzz uses Neural Networks to guide an evolutionary search over inputs of autonomous vehicle API~\cite{zhong2022neural}. The technique involves training the network to forecast whether new seeds will lead to unique traffic violations, with the most promising seeds being mutated to generate new adversarial inputs.

Our proposed method enhances test generation techniques by pinpointing the characteristics that have a significant impact on the safety-criticality of autonomous vehicles (AVs). This approach reduces the search space's complexity by narrowing down the features that require testing.

\subsection{Test Selection and Prioritisation}
To reduce the execution cost associated with AV testing, various test prioritisation and selection approaches are proposed. 
Arrieta et al. propose a search-based test-prioritisation method for cyber-physical systems product lines by incorporating five objectives in the fitness function i.e., fault detection capability, functional requirements covering time, non-functional requirements covering time, simulation time, and test case execution time~\cite{arrieta2019search}. In another study, Arrieta et al. propose a test selection technique for simulation models, using Non-Dominated Sorting Genetic Algorithm-II (NSGA-II), as a multi-objective search algorithm~\cite{arrieta2018multi}. In~\cite{ben2016testing}, multi-objective search is used to find safety-critical scenarios in \textit{Pedestrian Detection Vision System} for autonomous cars. The metric to guide the search is the distance between car and pedestrian, the distance between pedestrian and acute warning area, and time to collision (TTC)~\cite{vogel2003comparison}. In~\cite{lu2021search}, Birchler et al. propose to use static road features to prioritise the test cases for regression testing~\cite{birchler2021automated}. The search is guided by the fitness function that gives higher priority to the test cases having higher diversity and lower execution cost, which are estimated based on the previous runs of the test cases. Lu et al.~\cite{lu2021search} propose a test prioritisation technique for regression testing. Based on attributes of test cases e.g., speed, throttle, weather conditions etc., they compute four properties of test cases i.e., diversity, demand, collision probability and collision information, which are then used to guide the search to find the scenarios that lead to a collision. 
Deng et al. propose a test reduction and prioritisation approach for multi-module automated driving systems called
STRaP (Scenario-based Test Reduction and Prioritisation)~\cite{deng2022scenario}. Their approach encodes driving recordings into feature vectors using a driving scene schema and slices the recording into segments based on the similarity of consecutive vectors. They then truncate lengthy segments, remove redundant segments with the same vector, and prioritise the remaining segments based on both the coverage and rarity of driving scenes. It's worth noting that the feature vector employed in their approach bears a striking resemblance to the feature set utilised in our study. However, there are notable distinctions between our work and STRaP. Specifically, STRaP employs these features to identify stagnant recording segments and subsequently pinpoint and prioritise test scenarios that are capable of revealing failures during regression testing. In contrast, our approach, ISA, focuses on identifying features that exert a comprehensive influence on test outcomes. We employ these influential features to project instances into a two-dimensional (2D) space, enabling us to gain visual insights into the overall quality of the test suite.

The goal of the test prioritisation and selection techniques mentioned above is to cut the cost of the testing process by decreasing the size of the test suite. These techniques are proposed for regression testing in particular, to confirm that the recent code has not adversely affected the older features. Unlike these techniques, we propose to decrease the complexity of the search process by identifying the most impactful features challenging the safety of AV. The selected set of features can be used in test generation, prioritisation and selection. As the test scenarios for AV are composed of features to represent the environment, roads, AV, NPCs etc., some of the test generation and prioritisation techniques available in literature indirectly uses these features to compute fitness metric to guide search(e.g. diversity in ~\cite{lu2021search, birchler2021automated}). However, these techniques do not aim at finding an optimal set of features, and the diversity is computed based on the full feature vector used to design the test cases. In~\cite{birchler2021automated}, although a subset of features is selected from the complete road features using PCA, the goal of feature selection is to remove the correlated features without considering the impact of these features on test outcome (safe/unsafe). Furthermore, all the above-mentioned studies use test outcomes, like, execution time, time to collision, diversity of the test cases etc. as heuristics to guide the search towards safety-critical scenarios. Unlike these studies, we are interested in finding the compositional properties or input features of the test cases that should be considered to generate effective scenarios. 

\subsection{Test Scenarios Input Features}
Kluck et al.~\cite{kluck2021analysing} study the minimum number of features required to generate a test case that would lead to a collision. They consider two types of driving scenarios in their study: car-to-car and car-to-pedestrian scenarios generated in their previous work~\cite{kluck2019performance, kluck2019genetic}. Test cases are characterised by the properties of pedestrian members (start speed, offset, rate), AV (start speed, rate, target speed) and other participating vehicles (start speed, offset, target speed and rate). The number of vehicles ranges from 1 to 3 while the number of pedestrians ranges from 0 to 2. Decision trees are used to extract rules that lead to a crash either with pedestrians or other vehicles with high probability. It is shown that the combination of at least 4 parameters is required to generate a critical scenario with a high crash probability. Unlike~\cite{kluck2021analysing}, which finds the minimum number of features in a combination to make a scenario critical, we are interested in ``which'' features should be used and the range of their values (low or high), along with the possible combinations of these features that would increase the probability of a crash in a scenario. 

Cara and de Gelder use machine learning classifiers to classify car-cyclist scenarios into oncoming, longitudinal, or crossing scenarios using the trajectory features of the car and the cyclist \cite{cara2015classification}. Bircher et al. classify driving scenarios into safe and unsafe based on road static features~\cite{birchler2022cost}. In~\cite{kruber2018unsupervised, kruber2019unsupervised, kerber2020clustering, hauer2020clustering}, clustering techniques have been used to cluster traffic scenarios based on various features and distance measures. All of these techniques use compositional features of test scenarios for classification/clustering, however, finding the impact of features on test criticality has not been a research focus in these studies. Furthermore, although we have reported the prediction performance of four machine learning classifiers that classify the test-scenarios as safe or unsafe based on the feature values selected by ISA, the goal of feature selection in our proposed technique is multifold. ISA selects a subset of features that best defines a test scenario and clearly distinguishes a safe from an unsafe scenario, both in $nD$ feature space and $2D$ instance space. The instance space is created by making the projections such that the low values of features/test outcomes lie at one end of a straight line and high values at the other, i.e, linear trends of distribution of features and test result. Therefore, the impact of each feature on the performance can be clearly visualised, promoting the explainable feature selection. Furthermore, the instance space provides valuable insights into the coverage of the feature space and the diversity of the failed scenarios. 

\subsection{Test Adequacy Criteria}
To measure the sufficiency of system-level testing, various black-box coverage criteria and diversity metrics have been proposed~\cite{aghababaeyan2021black,hauer2019did,arcaini2021targeting,tang2021collision,majzik2019towards,laurent2022parameter}. Aghababaeyan et al. perform a correlation analysis of three black-box diversity measures (Geometric Diversity~\cite{kulesza2012determinantal}, Normalised Compression Distance~\cite{cohen2014normalized}, and Standard Deviation) with fault detection, and report that \textit{Geometric Diversity} shows the highest and statistically significant correlation. They further found that white-box coverage metrics are not an adequate measure for the quality of Deep Neural Networks (DNNs)~\cite{aghababaeyan2021black}. Hauer et al. propose to use ``types of test scenarios'' included in a test suite as a measure of coverage~\cite{hauer2019did}. Similarly, Arcaini et al. introduce some lower-level driving characteristics, such as ``turning with high lateral acceleration", and use them as a measure of coverage~\cite{arcaini2021targeting}. Tang et al. classify the scenarios based on the topological structure of the map and reported the performance of their proposed technique based on coverage of these structures~\cite{tang2021collision}. Majzik et al. provide a system-level situation coverage criteria with respect to the safety concepts captured by domain experts~\cite{majzik2019towards}. In~\cite{laurent2022parameter}, a parameter coverage criterion for Autonomous Driving Systems(ADSs) is proposed. Unlike these coverage measures, the coverage metric proposed in this study measures how extensively the feature space has been tested. It also identifies the critical regions where the probability of generation of safety-critical scenarios is high and thus should be prioritised to be tested.

Riccio et al. propose a search-based tool, DeepJanus, to find the pair of closely related test scenarios which shows different test outcomes (safe vs. unsafe), thus defining the frontier of behaviours of the Deep Learning (DL) system ~\cite{riccio2020model}. To assess the quality of the system, they introduce the notion of ``radius", which measures the distance between a nominal input (that the system is expected to handle correctly) and the input at the outer frontier. A high-quality system (HQ) shows a larger radius as compared to a low-quality (LQ), depicting that the system can tolerate larger changes to the input before exhibiting misbehaviour.

Jahangirova et al. extracted a set of 126 metrics that define the quality of human driving ~\cite{jahangirova2021quality}. A subset of 26 metrics is selected by a user study that aim to find the correlation between these metrics and human perception of driving quality of AVs. They further used these metrics as an oracle to assess the behaviour of an AV. 

A closely related work by Zohdinasab et al.~\cite{zohdinasab2021deephyperion} proposes a framework, known as DeepHyperion, which analyses the impact of features on the quality of the Deep Learning (DL) systems using illumination search~\cite{mouret2015illuminating}. The test instances are evaluated based on the fitness function, which measures the misbehaviour of the system-under-test, and the values for each test case are visualised in cells of a $2D$ map whose coordinates are defined by the two features of interest. The combination of the features that map to the cell having a maximum number of misbehaving instances (based on fitness value), is considered to have an impact on the misbehaviour. The map also provides a measure of the diversity of the failure-inducing inputs for feature combinations. A limitation of DeepHyperion is that it uses \textit{open coding procedure} to give a label to the input value of each feature, which is done manually. This procedure is done for every feature, which limits the scalability of the system and makes it infeasible to test instances having many features. Similar to DeepHyperion, ISA identifies the features having an impact on test outcomes and provides a quantitative measure of the coverage of the feature space. ISA, however, performs this process automatically and creates a footprint map of all features simultaneously, thus allowing the analysis of the relative importance of the different features.

\section{Threats to Validity}\label{sec:threats}

\textbf{Internal validity threats} exist when the presented results are influenced by internal factors. One such factor is the choice of features used in the current study. In order to mitigate this threat, we selected a wide range of features providing a detailed representation of AV, surrounding vehicles, pedestrians, road, and weather conditions. The representation of AV testing scenarios using some/all of these features is a common practice in AV-testing ~\cite{lu2021search,ding2022survey,li2020av, ebadi2021efficient}. 

Another factor could be using a different metric for the performance analysis of a testing scenario, such as \textit{collision probability}~\cite{lu2022learning}. This may result in the selection of different features, and thus the axis of the instance space might change, and results may differ. However, this does not impact the findings of the current paper, and the identified features significantly impact the effectiveness of test cases under the context that we have described in the experimental design section. 

\textbf{Conclusion threats to validity} pertain to issues that can affect the accuracy and reliability of the study's conclusions. One such threat in our study revolves around the legitimacy of feature combinations utilised in the boundary estimation process for coverage calculation. These combinations are deduced from test scenarios through feature correlation analysis. However, if the test suite contains scenarios that are either invalid or unrealistic, it can lead to the acquisition of incorrect correlations, thereby invalidating the boundary.

To mitigate this particular threat, we have taken a proactive approach by leveraging test suites from three prior studies that have been reported to encompass valid and realistic driving scenarios~\cite{lu2021search, frenetic, humeniuk2022ambiegen}.

\textbf{External validity threats} impact the generalizability of the results. To minimise these impacts, the data we used in our experiments is generated by testing Baidu Apollo -- an industrial-grade level-4 platform for AVs, and SVL-- an end-to-end autonomous vehicle simulation platform, and BeamNG.tech simulator. All of these tools are widely used in the literature of AV testing~\cite{ebadi2021efficient, li2020av, nguyen2021salvo, almanee2021scenorita, lu2021search, birchler2022single, birchler2021automated}. 

\section{Discussion}\label{sec:discussion}

The coverage metric, denoted as $\texttt{Coverage}_{IS}$, serves as a means to gauge the extent of coverage achieved by a test suite in relation to the extended instance space (boundary). This metric offers a visual representation of the coverage landscape, allowing testers to visually inspect areas that have been well covered and those that remain unexplored. This visualisation proves invaluable in identifying testing gaps and guiding efforts to enhance coverage by focusing on specific regions within the feature space.

The accuracy of $\texttt{Coverage}_{IS}$ hinges on the feasibility of boundary estimation, which entails defining the upper and lower limits of the features associated with driving scenarios. In our present study, the boundary calculation relies on the maximum and minimum feature values available within the test suite itself. Consequently, unoccupied regions in Figures~\ref{fig:boundaries_ds1} and~\ref{fig:boundaries_ds2} indicate scenarios absent relative to the test suite's content.

However, it's worth highlighting that ongoing research efforts, such as those related to Operational Design Domains (ODDs)~\cite{ye2022operational}, are actively exploring the establishment of absolute feature boundaries. These ODDs delineate the feasible driving conditions in which autonomous vehicles (AVs) can theoretically and practically operate. While initial progress has been made in defining these boundaries~\cite{ODDs}, coverage based on absolute boundaries would provide a measure of how comprehensively the entire feature space, as defined by potential driving conditions, is tested. This assessment would remain independent of the specific test suite used to generate the instance space.

\section{Conclusion and Future Work}\label{sec:conclusion}
This work presents a technique for the identification of input features of test scenarios that impact their effectiveness in terms of safety-criticality using Instance Space Analysis. ISA identifies features that show a high impact on the test outcome and generates a $2D$ instance space of test scenarios using these features. The distribution of feature values and test outcomes in the generated space helps to visually examine the impact of an individual feature and feature combinations on the test outcome. Furthermore, using the minimum and maximum values of feasible feature combinations, an area surrounding the instance space is identified where test scenarios can exist empirically, however, missing from the current instance space, giving an idea about the coverage of the current test suite.

ISA further trains five machine learning classifiers that classify the unlabelled test scenarios as safe or unsafe based on the selected features. The high performance of the classifiers, measured in terms of precision, recall, and f1-score, demonstrates that the selected features are predictive of the test scenario outcome and can be used to find the test result without simulating it, thus can help reduce the testing time. 

In our future work, we plan to develop a technique to fill the empty regions of the instance space by generating test scenarios having controlled properties. 

\section{DATA AVAILABILITY}
Test scenarios, metadata containing features and test outcome, and feature extraction code are publicly available at \url{https://github.com/neelofarhassan/ISA-for-AVs/tree/master}. The code for ISA is available at~\cite{isacode}.


\bibliographystyle{ACM-Reference-Format}
\bibliography{references}
\clearpage
\begin{appendix}
\section*{Appendices}
\section{Test Scenario: Test Suite 1}\label{sec:appendixa}
\begin{lstlisting}[language=json,firstnumber=1]

{
  "TimeStep1": {
    "Ego_Position": {
      "vx": 1.28,
      "ay": -0.48,
      "az": 9.81,
      "ax": -0.37,
      "pz": 10.13,
      "vy": 1.28,
      "px": 552874.26,
      "py": 4182784.82
    },
    "Ego_Operation": "SpeedCut",
    "Ego_Speed": "Slow (0.01 < speed (m/s) <= 5)",
    "Weather[rain]": "Light rain (0<rain_level<=0.2)",
    "Weather[fog]": "None fog (fog_level==0)",
    "Weather[wetness]": "Light wetness (0<wetness_level<=0.2)",
    "TimeofDay": "Noon (12pm)",
    "NPC": "None",
    "Pedestrian": "None",
    "Static obstacle": "None",
    "TrafficRule[Traffic light]": "Green (Allow to pass but slow at intersection)",
    "TrafficRule[Sidewalk]": "None",
    "CollisionInfoAtTimeStep": "NotOccurred"
  },
  "TimeStep2": {
    "Ego_Position": {
      "vx": 4.95,
      "ay": -0.75,
      "az": 9.81,
      "ax": -0.57,
      "pz": 10.13,
      "vy": 4.95,
      "px": 552882.81,
      "py": 4182778.27
    },
    "Ego_Operation": "EmergencyBrake",
    "Ego_Speed": "Fast (speed (m/s) > 8)",
    "Weather[rain]": "Light rain (0<rain_level<=0.2)",
    "Weather[fog]": "None fog (fog_level==0)",
    "Weather[wetness]": "Light wetness (0<wetness_level<=0.2)",
    "TimeofDay": "Noon (12pm)",
    "NPC": {
      "NPC1": {
        "position": {
          "y": 4182794.42,
          "x": 552891.13,
          "z": 10.34
        },
        "volume": "large",
        "operation": "SwitchLane (RightToLeft)",
        "speed": "Stop (0 < speed (m/s) <= 0.01)",
        "distance_temp": 18.17,
        "relativeDistance": "Far(18<distance<=28)"
      }
    },
    "Pedestrian": "None",
    "Static obstacle": "None",
    "TrafficRule[Traffic light]": "Green (Allow to pass but slow at intersection)",
    "TrafficRule[Sidewalk]": "None",
    "CollisionInfoAtTimeStep": "NotOccurred"
  },
  "TimeStep3": {
    "Ego_Position": {
      "vx": 4.45,
      "ay": -2.28,
      "az": 9.84,
      "ax": -1.75,
      "pz": 10.13,
      "vy": 4.45,
      "px": 552880.72,
      "py": 4182779.87
    },
    "Ego_Operation": "EmergencyBrake",
    "Ego_Speed": "Moderate (5 < speed (m/s) <= 8)",
    "Weather[rain]": "Light rain (0<rain_level<=0.2)",
    "Weather[fog]": "None fog (fog_level==0)",
    "Weather[wetness]": "Light wetness (0<wetness_level<=0.2)",
    "TimeofDay": "Noon (12pm)",
    "NPC": {
      "NPC1": {
        "position": {
          "y": 4182794.38,
          "x": 552891.17,
          "z": 10.28
        },
        "volume": "large",
        "operation": "SwitchLane (RightToLeft)",
        "speed": "Stop (0 < speed (m/s) <= 0.01)",
        "distance_temp": 17.88,
        "relativeDistance": "Near(8<distance<=18)"
      }
    },
    "Pedestrian": "None",
    "Static obstacle": "None",
    "TrafficRule[Traffic light]": "Yellow (Stop for a while)",
    "TrafficRule[Sidewalk]": "None",
    "CollisionInfoAtTimeStep": "NotOccurred"
  }
}
\end{lstlisting}
\section{Test Scenario: Test Suite 2}\label{sec:appendixb}
\begin{lstlisting}[language=json,firstnumber=1]
    {
  "test_id": "C:\\Users\\nnee0002\\Documents\\SDC-Scissors\\test-ambiegen\\test.0001.json",
  "is_valid": true,
  "test_outcome": "FAIL",
  "predicted_test_outcome": null,
  "test_duration": 3.875001907348633,
  "road_points": [
    [
      100.0,
      100.0
    ],
    [
      99.927,
      100.998
    ],
    [
      99.825,
      101.993
    ],
    [
      99.695,
      102.985
    ],
    [
      99.54,
      103.973
    ]
    .
    .
    .
  ],
  "interpolated_road_points": [
    [
      99.99999999999999,
      99.99999999999997,
      -28,
      10
    ],
    [
      99.5265443923853,
      104.0518028306302,
      -28,
      10
    ],
    [
      98.66455037194862,
      108.04008039802935,
      -28,
      10
    ],
    [
      97.55097089076908,
      111.96627024021502,
      -28,
      10
    ],
    [
      96.32578599276616,
      115.85940468585808,
      -28,
      10
    ]
    .
    .
    .
]
}
\end{lstlisting}
\end{appendix}
\end{document}